# An Optimized Multi-Layer Resource Management in Mobile Edge Computing Networks: A Joint Computation Offloading and Caching Solution

Amir Ziaeddini[1], Amin Mohajer[1], Davoud Yousefi[2], A.Mirzaei[2], Shu Gonglee[3*]
[1] Mobile telecommunication company of Iran (MCI), Tehran, Iran
[2]Department of Computer Engineering, Ardabil Branch, Islamic Azad University, Ardabil, Iran
[3]Department of Computer Science, Chengdu University of Technology, Sichuan 610059, China

[*]Corresponding Author (*email address: shu.g.gust@gmail.com*)

**ABSTARCT**
Nowadays, data caching is being used as a high-speed data storage layer in mobile edge computing networks employing flow control methodologies at an exponential rate. This study shows how to discover the best architecture for backhaul networks with caching capability using a distributed offloading technique. This article used a continuous power flow analysis to achieve the optimum load constraints, wherein the power of macro base stations with various caching capacities is supplied by either an intelligent grid network or renewable energy systems. This work proposes ubiquitous connectivity between users at the cell edge and offloading the macro cells so as to provide features the macro cell itself cannot cope with, such as extreme changes in the required user data rate and energy efficiency. The offloading framework is then reformed into a neural weighted framework that considers convergence and Lyapunov instability requirements of mobile-edge computing under Karush–Kuhn–Tucker optimization restrictions in order to get accurate solutions. The cell-layer performance is analyzed in the boundary and in the center point of the cells. The analytical and simulation results show that the suggested method outperforms other energy-saving techniques. Also, compared to other solutions studied in the literature, the proposed approach shows a two to three times increase in both the throughput of the cell edge users and the aggregate throughput per cluster.

## 1. INTRODUCTION

Next-generation cellular networks require up to about a thousand times more data rate capacity than current long-term networks [1, 2, 3]. In view of its benefits, sub-channel allocation is adapted using the non-orthogonal frequency division multiple access (NOMA) technique [4, 5, 6]. In such networks, the first challenge is to obtain the Channel State Information (CSI) of the co-channel users. These reliable communications face to strong residual interference among NOMA users without the knowledge of CSI [7, 8, 9, 10]. The second challenge is to remove the interference from the different signals of the co-channel users. A novel interference cancellation technique is also necessary to feat information of and suppresses the interference from all the overlapping symbols of the cochannel interference. It is accomplished using spectrum reuse techniques with NOMA known as the user-pairing algorithm. The intra-cell interference is reduced and limited by using spectrum reuse techniques with Hybrid NOMA known as the user-pairing algorithm [11, 12, 13, 14, 15].

On the other hand, SLA-based services, which need low-latency communications, currently account for more than 60% of all data services, according to ITU technical publications [16]. Given the increasing demand for high-quality services, resource allocation and load sharing techniques in mobile edge computing networks (multi-layer network architecture) are becoming a major issue [17, 18, 19, 20]. To address these concerns, data offloading between core networks and other levels of the hierarchical network might be regarded as an efficient means of reducing network latency and high usage challenges. Earlier studies, such as [21, 22, 23], analyzed the effectiveness of next-generation cellular networks using new backhauling routing techniques. In [24, 25], the authors concentrated on caching-based mobile edge computing.

In order to meet the SLA constraints, and to determine the ideal setup of MEC platforms, [26, 27, 28] propose a mixed-integer quadratic function limited joint allocation of resources using linear programming, which is evaluated using multiple trustworthy NOMA scenarios. Various meta-heuristic techniques with diverse target modules are used in the reconfiguration cycle. In [29], the evolutionary algorithm is frequently used to find the most effective switching solution using many different kinds of neural networks. Several meta-heuristic techniques using safety optimization problems are used in the remodeling phase [30, 31]. To solve these restructuring challenges, particle swarm optimization (PSO) [32], harmony search (HS) [33], and ant colony techniques [34] have been applied. In [35], the researchers suggested a cache-enabled queue strategy for HetNet systems, concentrating on the problem of optimum data transmission, which they characterized as a stochastic optimization model. This was demonstrated in [36], suggesting employing a multihop relay strategy can enhance caching performance. It's also suggested in [37, 38] that the small cells may be grouped into clusters and utilized in combination with their caches to hold storage efficiency. The researchers [39] addressed the issue of establishing a shared cache and assigning caching capacity in mobile device-to-device networks in order to optimize cache utilization. 5G networks, which are expected to be among the most important platforms for implementing this critical technology, are expected to be significantly more energy-efficient. Energy efficiency with renewable energy sources is an appealing way to boost energy efficiency. Certain base stations with high power consumption cannot entirely cache the requested data, while others may not get enough energy to transfer data. Energy collaboration has already been suggested to alleviate the problem of load balance, taking into account changes in traditional networks and improvements in energy flow. As a result, base stations may offer more power for stations that are experiencing power shortages. Certain power control strategies and energy cooperation initiatives have indeed been proposed in certain studies.

The writers of [40] explored the subject of power-sharing across micro-grids. [41] Studied allocation of resources and power-sharing to improve the EE indices in HetNets. Earlier research [42, 43] established the framework for power savings in caching-based and energy cooperation systems. Nonetheless, no work has been done on resource allocation in energy-cooperative systems with shared cache capabilities, and it remains an unexplored field of study. In light of something like this, the proposed method proposes an optimal paradigm for UA and resource allocation in caching-based power-cooperative HetNets. The goal is to boost total system performance while lowering traditional grid power use. Cache-enabled and energy-cooperative MEC networks are assumed to contain macro base stations (MBS) and Small base stations (PBS). Each one includes a cache for holding contents files. Suppose B and U represent the base station and user equipment (UE), respectively. Each macro and Small base station has a cache capacity of $L_M$ and $L_S$, respectively.

This study looked at the influence of efficient resource allocation in mobile edge computing networks as well as the data flow of the available connections in backhauling networks. We will look at two ways in this work, taking into account the rigorous limitations on capacity requests by all user equipment (UE) and authorizing the transmission of a suitable request to the user's equipment. This minor adjustment seems to have a big impact on how both challenges are phrased. Because the suggested methodology is convex and quasi-convex, a bi-section method-based technique was recommended for finding the best solution. The quantitative findings demonstrate the efficacy of the suggested technique in comparison to previous baseline EE methods.

The next generation MEC networks are intended to boost data speed and bandwidth, give Internet coverage and connection everywhere, and dramatically reduce power consumption [44, 45, 46, 47]. Recent studies in the field of 5G has concentrated on multi-layered hierarchical allocating resources and optimizations, that are crucial to successful innovations [48, 49, 50]. Base stations with various transmit powers, coverage areas, carrier frequencies, and kinds of BH (backhaul) connections were employed in HetNets, which introduces additional issues in linking all base stations to the core network through adequate backhaul links [51]. As a result, before backhaul traffic gets to the main network, the base station might well be forced to share the link with other core network nodes.

## 2. NETWORK MODEL

Each base station in the MEC energy cooperative networks receives power from either microgrid supplies or renewable power systems. The weighting coefficient *l* is utilized as the time factor during the procedure to reduce computing complexity. The transmission energy of base station *i* is $P_i(i \in B)$, the used power from the traditional grid is $G_i$, and the base station $i$ can collect special energy from renewable sources power systems is $E_i$.

Base station $i$ may share its energy with base station i equal to $E_i$, using an $\beta\epsilon[0,1]$ for the power distribution. Since the power usage time frame might be considerably longer than the process of power savings [52], it's often assumed that no energy is stored. As a result, the continuity equation for the transmission rate of the ith base station should be developed. As a result, the continuity equation for the transmission rate of the ith base station should be developed. In the suggested cache-enabled MEC network's system model, each cellular cluster has one macro cell and multiple small cells, where U and B denote the set of user equipment and base stations, respectively. LS and LM denote the cache size of each small base station and macro base station, respectively, in which the energy for each cell can come from both renewable and conventional sources. It should be noted that the base stations in this energy cooperative network are willing to share their resources via smart grids. The system setup of this cache-enabled network has been presented in Fig. 1 according to the specifications. In this article, it is expected that numerous mixed backhaul connections between Small base stations (PBS) and one backhaul connection between the cluster head and the macro base station are accessible. Without deviating from the main point, it is believed that the macro base station (MBS) is linked to the central network through a fiber link. Each small base station is connected directly to the macro base station via a cluster head or one or more small base station density points.

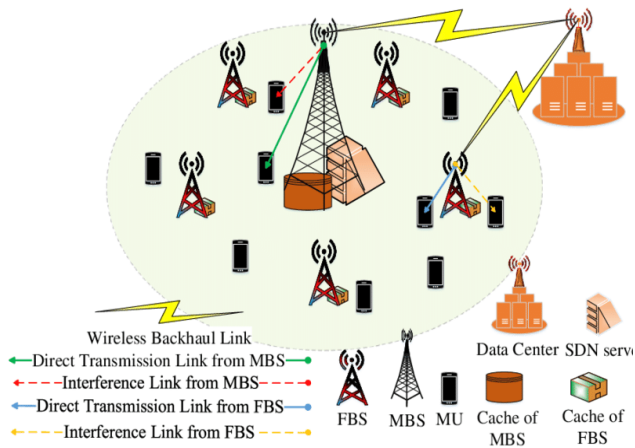

**Fig.1.** MEC network configuration with caching capability

In a backhaul system, two sub-carrier frequencies could be employed. The article utilized 73 GHz backhaul links (band E) and 60 GHz band (V band) for numerous backhaul links between small base stations for communication. Millimeter-wave transmission dissipation ratios are divided into two main types: path loss and millimeter-wave transmission dissipation factors. These criteria were clearly indicated in Figure 2.

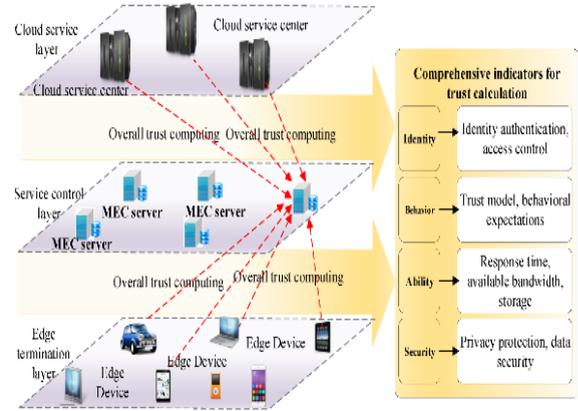

**Fig.2.** Multi-layer configuration of the cache-enabled mobile edge computing

According to the leading publications in this field of study, there is no standard model for energy usage in MEC backhaul networks. Nevertheless, the implementation of nonlinear forecast energy demand in millimeter waves has received the most attention. This research employs adaptive resource allocation to simulate the backhaul link, where C5 and C6 are the highest transmission power restrictions for small and macro base stations, respectively.

## 2.1 CONTENT‑ CACHING MODEL

We assume in this network that information may be represented as a discrete collection of data packets as $F = \{F_1, F_2, \ldots, F_f, \ldots, F_F\}$, where $F_f$ denotes the f-th data frame. The demand possibility for data frame $f$ is written as follows:

$$p_f(0 \leq p_f \leq 1), \text{ which}, \sum_{f=1}^{F} p_f \leq L_i, \ \forall f \in \mathcal{F}, \quad (1)$$

It should have been mentioned that the caching method proposed in this study is probabilistic caching, which

16

allows us to determine the possibility of caching data packet $f$ through base station i as $0 \leq q_{f_i} \leq 1$, where $L_i$ represents the cache size. Furthermore, the $\{q_{f_i}\}$ of base station $i$ must meet the following criteria:

$$\sum_{f=1}^{F} q_{f_i} \leq L_i, \quad \forall i \in \mathcal{B}, f \in \mathcal{F}, \tag{2}$$

## 2.2 Resource Control Model

According to the approach's concepts, the base stations' resources may be provided by traditional smart grid and renewable energy collection. Throughout this situation, $P_i (i \in B)$ represents the transmit power relevant to base station $i$ and $G_i$ represents the supplied energy from the grid network. $E_i$ represents the collected renewable sources. Depending on the activated energy capabilities, the shared power between cells i and $i'$ equals $\varepsilon_{ii'}$, where $\beta \in [0,1]$ signifies the power-sharing indices across base stations. As a result, we may deduce that $(1 - \beta)$ equals the loss ratio at the power-sharing stage. During the power-sharing procedure, the approaches were adopted.

$$P_i < G_i + E_i + \beta \sum_{i' \in \mathcal{B}, i' \neq i} \varepsilon_{i'i} - \sum_{i' \in \mathcal{B}, i' \neq i} \varepsilon_{ii'}. \tag{3}$$

Transmission systems, power-sharing, and the amount of captured energy from renewable resources can all have an impact on total power savings, depending on the severity.

## 2.3 Transmission Model

We attempted to offer data rate balancing throughout the network while keeping fairness in mind. For example, $x_{ij} = 1$ indicates that UE j is connected to BS i, whereas $x_{ij}(i \in B, j \in u)$ indicates that the user is not connected to the base station. Following that, $k_i$ denotes the number of users linked with cell i. So that, $k_i = \sum_{f=1}^{F} p_f q_{f_i}$ denotes the number of users associated with cell i. $\left(\sum_{f=1}^{F} p_f q_{fi}\right)^{k_i}$ denotes the possibility of base station I supporting $k_i$ connected UEs. If $x_{ij} = 1$, we can compute the efficacy of the j-th user equipment as $\mu_{ij} = log(R_{ij})$ where $R_{ij}$ is the user output and the $R_{ij}$ is attainable as.

$$R_{ij} = \left(\sum_{f=1}^{F} p_f q_{f_i}\right)^{k_i} \frac{\mathcal{B}}{\sum_{j \in u} x_{ij}} \log(1 + \gamma_{ij}) \tag{4}$$

The signal-to-noise ratio may be calculated using this approach (5).

$$\gamma_{ij} = \frac{P_i h_{ij}}{\sum_{i' \in \mathcal{B}, i' \neq i} P_{i'} h_{i'j} + \sigma^2} \tag{5}$$

In this equation, $h_{ij}$ and $h_{ij'}$ signify the main channel gains and the interference channel gains, respectively, while B specifies the frequency bandwidth. $\sigma^2$ is also the "noisy" figure. The objective function is similar to minimizing the applied grid power. As a result, the aim function is as follows:

$$\textbf{P1:} \quad \max_{q,x,P,\varepsilon,G} \sum_{i \in \mathcal{B}} \sum_{j \in u} x_{ij}\mu_{ij} - \eta \sum_{i \in \mathcal{B}} G_i \tag{6}$$

s. t.
$$C1: \sum_{i \in \mathcal{B}} x_{ij}\gamma_{ij} \geq \gamma_{min}, \forall j \in u,$$

$$C2: \sum_{i \in \mathcal{B}} x_{jm} = 1, \forall j \in u,$$
$$C3: P_i < G_i + \beta \sum_{i' \in \mathcal{B}, i' \neq i} \varepsilon_{i'i} - \sum_{i' \in \mathcal{B}, i' \neq i} \varepsilon_{ii'} + E_i, \forall i \in \mathcal{B},$$
$$C4: \sum_{f=1}^{F} qf_i \leq L_i, \forall i \in \mathcal{B}, f \in \mathcal{F},$$
$$C5: 0 \leq q_{f_i} \leq 1, \forall f \in \mathcal{F}, \forall i \in \mathcal{B},$$
$$C6: x_{ij} \in \{0,1\}, \forall i, \forall j \in u,$$
$$C7: G_i \geq 0, \varepsilon_{ii'} \geq 0, \forall i \in \mathcal{B},$$
$$C8: 0 \leq P_i \leq P_{max}^i, \forall i \in \mathcal{B},$$

In this system model, $q = [q_{f_i}], x = [x_{ij}], P = [p_i], \varepsilon = [\varepsilon_{ij'}], G = [G_i], \gamma_{min}$ represents $SNIR_{min}$ to ensure the dependability of the connection between the UEs and the base station. And $\eta$ is a weighting factor for calculating power usage. The multihop backhauling approach and network architecture of the described MEC network are shown in Fig. 3.

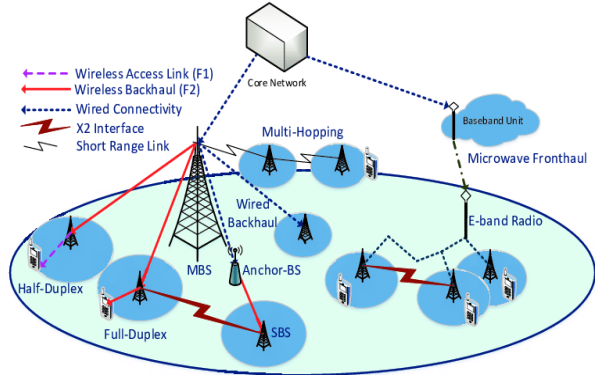

**Fig.3.** multi-hop backhauling concept in the MEC architectural style.

## 2.4 CONTINUATION POWER FLOW (CPF) AND OPTIMIZATION

To address the aforementioned issues, optimizing energy efficiency (EE) in HetNets has already been examined independently in the fields of access networks [53] and backhaul networks [54], as well as in mixtures. Continuation power flow as a method for tackling the Jacobian matrix issue in SLA-based flow equations, getting optimum at the power steady state by discovering frequent load flow solutions in power-constrained contexts [55].

There are two phases to this method. Stages of forecasting and corrections a tangential indication is utilized in this approach to get the estimated second solutions from a BS for a special load increase. The corrector step, which employs the N-R approach of traditional power flow, likewise yields the best result. This procedure will be repeated until the tangential solution equals zero, corresponds to the vital point. Figure 1 depicts the predictor–corrector method.

The power input in the paradigm of the continuing power flow for the ith of a *n* bus system could be written as:

$$P_i = \sum_{k=1}^{n} |V_i||V_K|(G_{ik}cos\theta_{ik} + B_{ik}sin\theta_{ik}) \quad (7)$$

$$Q_i = \sum_{k=1}^{n} |V_i||V_K|(G_{ik}sin\theta_{ik} + B_{ik}cos\theta_{ik}) \quad (8)$$

$$P_i = P_{Gi} - P_{Di}; \quad Q_i = Q_{Gi} - Q_{Di}; \quad (9)$$

In so that G and D represent generating and load requests, respectively. To predict load variation, the load factor $\lambda$ is used with actual and reactive resource requirements.

$$P_{Di} = P_{Dio} + \lambda(P_{\Delta base}) \quad (10)$$
$$Q_{Di} = Q_{Dio} + \lambda(P_{\Delta base}) \quad (11)$$

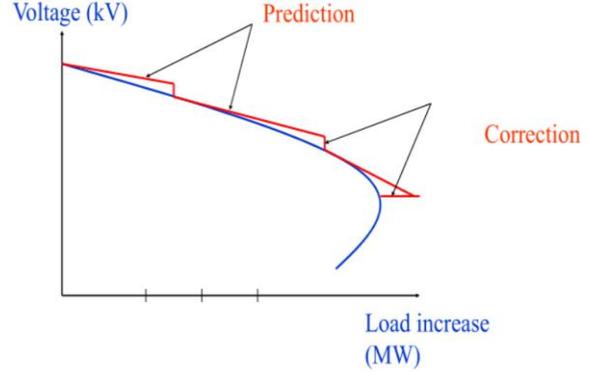

**Fig. 4.** continuous power flow numerical conceptualization

Considering $P_{Dio}$, we could visualize the actual load needs on the fifth bus. To measure approximately $\lambda Q_{Dio}$, $P_{\Delta base}$, and $Q_{\Delta base}$ were used. It is reasonable to view novel resource requirements in (9) to (11) new formulations:

$$F(\theta, V, \lambda) = 0 \quad (12)$$

The bus power ratios are denoted by $\theta$, while the bus power amplitudes are denoted by V. With a match of $\lambda = 0$, the main equation is solved.

(1) Foreseeability Stage: By selecting a suitable scaled stage in the tangential formula, a linear approximation is applied within that stage. Therefore, we have

$$F_\theta d_\theta + F_V d_V + F_\lambda d_\lambda = 0 \quad (13)$$

$$|F_\theta \quad F_V \quad F_\lambda| * \begin{vmatrix} d_\theta \\ d_v \\ d_\lambda \end{vmatrix} = 0 \quad (14)$$

Setting one of the tangential solution elements to +1 or 1 (the continuance variable) and imposing a quantity to the tangential vertex leads the Jacobian to be not unique at the vital moment.

$$\begin{vmatrix} F_\theta & F_V & F_\lambda \\ & e_k & \end{vmatrix} * \begin{vmatrix} d_\theta \\ d_v \\ d_\lambda \end{vmatrix} = \begin{vmatrix} 0 \\ \pm 1 \end{vmatrix} \quad (15)$$

Wherein $e_k$ designates the appropriate row vector with the $k^{th}$ element equal to one and all other elements equal to 0. Normally, $\lambda$ shows the continuation indicator.

18

Through finding solutions (15), the tangent result is found, and the forecast would be as follows:

$$\begin{vmatrix} \theta \\ V \\ \lambda \end{vmatrix}^{p+1} = \begin{vmatrix} \theta \\ V \\ \lambda \end{vmatrix}^{p} + \sigma \begin{vmatrix} d_\theta \\ d_V \\ d_\lambda \end{vmatrix} \quad (16)$$

The forecasting is signified by the symbol p + 1 in such formulas and $\sigma$ denotes the step length. At which point the expected outcome would be situated inside the corrector's convergence.

(2) Stage of Correction: As expected, the expected stage is accessible in the corrector phase via local generalization. The actual solution would be enlarged to the given value of the state variable, yielding the expected result.

$$\begin{vmatrix} F(\theta, V, \lambda) \\ x_k - \eta \end{vmatrix} = |0| \quad (17)$$

Wherein $x_k$ represents the state variable chosen as the continuation component and $\eta$ represents the expected state variable. In addition, the best outcome for (17) could be found using a personalized Newton–Raphson power flow technique.

## 3. PROPOSED USER ASSOCIATION AND MEC RESOURCE ALLOCATION

Considering the system model's restrictions, the issue of user association and power allocation can be modeled as problem P2. That is the best approach for the original problem P1. This issue may be phrased as follows when $k_i = \sum_{j \in u} x_{ij}$ is taken into consideration.

$$\text{P2:} \quad \max_{q,x,P,\varepsilon,G} \sum_{i \in \mathcal{B}} \sum_{j \in u} x_{ij} \log(c_{ij}) + \sum_{i \in \mathcal{B}} k_i^2 \log(\sum_{f=1}^{F} p_f q f_i) - \sum_{i \in \mathcal{B}} k_i \log(k_i) - \eta \sum_{i \in \mathcal{B}} G_i \quad (18)$$

s.t. C1, C2, C3, C4, C5, C6, C7, C8,

$$C9: \sum_{j \in u} x_{ij} = k_i, \forall i,$$

Where, $c_{ij} = B \log(1 + \gamma_{ij})$.

*3.1- User Association Algorithm Based on Data Caching*

$$\text{P2.1:} \quad \max_{q,x} \sum_{i \in \mathcal{B}} \sum_{j \in u} x_{ij} \log(c_{ij}) + \sum_{i \in \mathcal{B}} k_i^2 \log\left(\sum_{f=1}^{F} p_f q f_i\right) - \sum_{i \in \mathcal{B}} k_i \log(k_i) \quad (19)$$

s. t. C1, C2, C4, C5, C6, C9

Lemma 1: Given p(1) ≥ ... ≥ p(f) ≥ ... ≥ p(F) as the possibility of desired payload (f), the best course of action for P2.1 is as follows.

$$q^*_{f_i} = \begin{cases} 1, & f_i = (1), \dots, (L_i) \\ 0, & otherwise \end{cases}, \quad \forall i \in \mathcal{B}. \quad (20)$$

Theory 1: As previously stated, P2.1 demonstrates that achieving the optimum solution for $\sum_{f=1}^{F} p_f q_{fi}$ that the goal wherein the required payload contains $F(l = 1, \cdots, Li)$ segments and $F_l$ possibility is greater than $F_{l+1}$. As a result, we have:

$$\sum_{(f) \in \mathcal{F}_l} q^l_{(f)_i} = 1, \sum_{l=1}^{L_i} q^l_{(f)_i} = q_{(f)_i} \text{ and } \bigcup_{l=L_i} \mathcal{F}_l = \mathcal{F}$$

Also,

$$\sum_{f=1}^{F} p_f q_{fi} = \sum_{l=1}^{L_i} \sum_{(f) \in \mathcal{F}l} p_{(f)} q^l_{(f)i} \leq \sum_{l=1}^{L_i} p_{(l)} \left(\sum_{(f) \in \mathcal{F}l} q^l_{(f)i}\right) \Rightarrow \sum_{f=1}^{F} p_f q_{f_i} \leq \sum_{l=1}^{L_i} p_{(l)},$$

As a result, according to (20), this idea is supported and we may deduce that

$$\tilde{P}2.1: \quad \max_{x} \sum_{i \in \mathcal{B}} \sum_{j \in u} x_{ij} \log(c_{ij}) + \sum_{i \in \mathcal{B}} k_i^2 \log\left(\sum_{f=1}^{L_i} p_{(f)}\right) - \sum_{i \in \mathcal{B}} k_i \log(k_i) \quad (21)$$

s.t C1,C2,C6,C9.

We concentrate on both its issues to simplify the process of obtaining the best response to P 2.1 as a mixture of numerous sub-problems. As a result, the leveling process should be written in the following form:

$$\mathcal{L}(x,k,\mu,\nu) = \sum_{i \in \mathcal{B}} \sum_{j \in u} x_{ij} log(c_{ij}) + \sum_{i \in \mathcal{B}} k_i^2 log\left(\sum_{f=1}^{L_i} p_{(f)}\right) - \sum_{i \in \mathcal{B}} k_i log(k_i) - \sum_{j \in u} \mu_j \left(\gamma_{min} - \sum_{i \in \mathcal{B}} x_{ij} \gamma_{uj}\right) - \sum_{i \in \mathcal{B}} \nu_i \left(\sum_{i \in u} x_{ij} - k_i\right), \quad (22)$$

Wherein $v = [v_i], k = [k_i]^1$ and $\mu = [\mu_j]$ in this equation.

It's worth noting that $v_i, \mu_j$ are Lagrange factors. To proceed, we can write D(.) as the dual function of the problem.

$$\mathcal{D}(\mu,v) = \begin{cases} \max_{x,k} \mathcal{L}(x,k,\mu,v) \\ s.t. \; C2, C6. \end{cases} \quad (23)$$

The dual problem of P 2.1 (21) will therefore be reformulated as

$$\min_{\mu \geq 0, v \geq 0} \mathcal{D}(\mu,v). \quad (24)$$

The coefficients of the dual problem are $v_i$, $\mu_j$ and the goal function result may be determined by following these steps:

$$x_{ij}^* = \begin{cases} 1, & if \; i = i^* \\ 0, & otherwise \end{cases}, \quad (25)$$

$i^* = arg\,maxi\left(\log(c_{ij}) + \mu_j y_{ij} - v_i\right)$ is used in (25). In the case of $k_i$, the second derivation of the goal function yields

$$\frac{\partial^2 \mathcal{L}}{\partial k_i^2} = 2log\left(\sum_{f=1}^{L_i} p_{(f)}\right) - \frac{1}{k_i}. \quad (26)$$

$\sum_{f=1}^{L_i} P(f) \leq 1$ and $\frac{\partial \mathcal{L}}{\partial k_i^2}$ cannot be a positive number, since it is clear. $K_i^*$ is attained as the optimal level of $k$ when $\frac{\partial \mathcal{L}}{\partial k_i^2}$ is set to 0.

$$k_i^* = -\frac{W\left(-2log\left(\sum_{f=1}^{L_i} p_{(f)}\right)e^{v_i-1}\right)}{2log\left(\sum_{f=1}^{L_i} p_{(f)}\right)}, \quad (27)$$

$w(z)$ represents the Lambert-W factor as a reaction for $z = we^w$ in (27). According to (25), differentiating $(\mu,v)$ does not result in the best solution $(\mu^*, v^*)$. As a result, using the iterative gradients technique will be beneficial.

$$\mu_j(t+1) = \left[\mu_j(t) - \delta(t)\left(\sum_{i \in \mathcal{B}} x_{ij}(t)\gamma_{ij} - \gamma_{min}\right)\right]^+, \quad (28)$$

$$v_i(t+1) = \left[v_i(t) - \delta(t)\left(k_i(t) - \sum_{j \in u} x_{ij}(t)\right)\right]^+, \quad (29)$$

Using (25) and (27), $x_{ij}(t)$ and $k_i(t)$ can be refreshed in a repetition of this equation. $\delta(t)$ represent the step length, and $[a]^+ = max\{a, 0\}$ represents the number of iterations.

It's worth noting that this algorithm is certainly convergent since it meets all of the convergence criteria in [56].

## 4. CACHING-ENABLED OUTAGE PROBABILITY CONTROL

This paper proposes a novel approach to support a novel transmission strategy considering non-orthogonal multiple access for multi-server heterogeneous IoT networks.

One of the main critical issues in capacity management is control of outage probability. In order to derive a closed-form expression for the outage probability of the proposed UE selection and relay association coded system, we assume non-identically and independently distributed Rayleigh fading over all multi-server heterogeneous network connections.

In this model, if $X_1, X_2, ..., X_n$ is set of $n$ different independent random variable with PDF: $f_{x_i}(x_i) = \lambda_i e^{-\lambda_i x_i}$. So, we have,

$$Pr\{X_1 > X_2 > \cdots > X_n\} = \prod_{v=2}^{n}\left[\frac{\lambda_v}{\lambda_1 + \sum_{i=2}^{v}\lambda_i}\right] \quad (30)$$

We can rewrite the probability function (30) in the integral form as (31)

$$Pr\{X_1 > X_2 > \cdots > X_n\}$$
$$= \underbrace{\int_0^\infty \int_0^{x_1} \int_0^{x_2} \cdots \int_0^{x_{n-1}}}_{n} \prod_{i=1}^{n}[fx_i(x_i)]dx_n dx_{n-1} \ldots dx_1. \quad (31)$$

Considering (30), and $n = 2$, the outcome of (31) is obtained as

$$Pr\{X_1 > X_2\} = \int_0^\infty \int_0^{x_1} \lambda_1 e^{-\lambda_1 x_1} \lambda_2 e^{-\lambda_2 x_2} dx_2 dx_1$$
$$= \int_0^\infty \lambda_1 e^{-\lambda_1 x_1}(1 - e^{-\lambda_2 x_1})dx_1$$
$$= \frac{\lambda_2}{\lambda_1 + \lambda_2}, \quad (32)$$

We can also suppose that equation (30) holds when $n = k$.

$$\Pr\{X_1 > X_2 > \cdots > X_k\} = \prod_{v=2}^{k}\left[\frac{\lambda_v}{\lambda_1 + \sum_{i=2}^{v}\lambda_i}\right] \quad (33)$$

subsequently, it should be proved that (30) is correct when $n = k+1$ considering (33):

$$\Pr\{X_1 > X_2 > \cdots > X_{k+1}\} = \Pr\{X_1 > X_2 > \cdots > X_k\}\underbrace{\Pr\{X_{k+1} < X_1, X_2, \ldots, X_k\}}_{B}. \quad (34)$$

In (22) the probability $B$ can be expressed as the following

$$B = \underbrace{\int_0^\infty \int_{x_{k+1}}^\infty \int_{x_{k+1}}^\infty \cdots \int_{x_{k+1}}^\infty}_{k+1} \prod_{i=1}^{k+1}[f_{X_i}(x_i)]dx_1 \ldots dx_{k+1}$$
$$= \lambda_{k+1}\int_0^\infty e^{-(\lambda_1+\lambda_2+\cdots+\lambda_{k+1})x_{k+1}}dx_{k+1}$$
$$= \frac{\lambda_{k+1}}{\lambda_1 + \lambda_2 + \cdots + \lambda_{k+1}}. \quad (35)$$

Considering (22) and (23), we conclude:

$$\Pr\{X_1 > X_2 > \cdots > X_{k+1}\}$$
$$= \prod_{v=2}^{k}\left[\frac{\lambda_v}{\lambda_1 + \sum_{i=2}^{v}\lambda_i}\right]\left[\frac{\lambda_{k+1}}{\lambda_1 + \lambda_2 + \cdots + \lambda_{k+1}}\right]$$
$$= \prod_{v=2}^{k+1}\left[\frac{\lambda_v}{\lambda_1 + \sum_{i=2}^{v}\lambda_i}\right]. \quad (36)$$

For the special case of identically and independently distributed random variables such as, $\lambda_i = \lambda, \forall i$, (30) can be simplified as (37)

$$\Pr\{X_1 > X_2 > \cdots > X_n\} = \frac{1}{n!} \quad (37)$$

Based on the problem assumption of the scenario, if the final receiver selects the $i_1^{th}, i_2^{th}, \ldots i_K^{th}$, the best UEs and $j_1^{th}, j_2^{th}, \ldots j_L^{th}$, as the best intermediate relays, the system's close-form expression of the outage probability when $K > L$ will be achieved as (38)

$$\mathcal{P}_{out_1} = \sum_{\eta=0}^{K-L-1}\Pr\{\varepsilon_\eta\} + \sum_{\eta=1}^{L}\left(\Pr\{\varepsilon_{K-\eta}\}\sum_{\ell=0}^{\eta-1}\Pr\{\mathcal{V}_\ell\}\right) \quad (38)$$

Subsequently, for $K \leq L$ the outage probability of the system is obtained as

$$\mathcal{P}_{out_2} = \sum_{\eta=1}^{K}\left(\Pr\{\varepsilon_{K-\eta}\}\sum_{\ell=0}^{\eta-1}\Pr\{\mathcal{V}_\ell\}\right) \quad (39)$$

$$\Pr\{\varepsilon_\eta\} = \sum_{v=N-i_\eta+1}^{N-i_\eta}\left(\sum_{\substack{a_1,\ldots,a_v \in \{1,\ldots,N\} \\ a_1 \neq \cdots \neq a_v}}\left[\prod_{m=a_1}^{a_v}\Pr\{\mathcal{O}_{S_nD}\}\prod_{\substack{\dot{n}=1 \\ \dot{n}\neq\{a_1,\ldots,a_v\}}}^{N}(1 - \Pr\{\mathcal{O}_{S_{n'}D}\})\right]\right). \quad (40)$$

$\Pr\{\mathcal{V}_\ell\}$
$$= \sum_{\mathcal{A}}\left(\sum_{v=M-j_\ell+1}^{M-j_\ell}\left(\sum_{\substack{a_1,\ldots,a_v \in \{1,\ldots,M\} \\ a_1 \neq \cdots \neq a_v}}\left[\prod_{m=a_1}^{a_v}\Pr\{\mathcal{O}_m|\mathcal{A}\}\prod_{\substack{\dot{m}=1 \\ \dot{m}\neq\{a_1,\ldots,a_v\}}}^{M}(1 - \Pr\{\mathcal{O}_{\dot{m}}|\mathcal{A}\})\right]\right)\times \sum_{\substack{z_1,\ldots,z_N \in \{1,\ldots,N\} \\ z_1 \neq \cdots \neq z_N \\ z_{i_1},\ldots,z_{i_K} \in \mathcal{A}}}\left(\prod_{n=2}^{N}\left[\frac{\lambda_{S_{z_n}D}}{\lambda_{S_{z_1}D} + \sum_{i=2}^{n}\lambda_{S_{z_i}D}}\right]\right)\right). \quad (41)$$

In which $\Pr\{\varepsilon_\eta\}$ and $\Pr\{\mathcal{V}_\ell\}$ are, obtainable via (40) and (41). So, $\Pr\{\mathcal{O}_{S_nD}\}$ is calculated as (42)

$$\Pr\{\mathcal{O}_{S_nD}\} = 1 - e^{-\lambda_{S_nD}\gamma_{th}}, \quad (42)$$

With $\lambda_{ij} = \frac{1}{\rho\sigma_{ij}^2}$.

Also, $\Pr\{\mathcal{O}_m|\mathcal{A}\}$ is calculated by (43).

$$\Pr\{\mathcal{O}_m|\mathcal{A}\} = 1 - e^{-\lambda_{m|\mathcal{A}}\gamma_{th}}, \quad (43)$$

with $\lambda_{m|\mathcal{A}}$ being

$$\lambda_{m|\mathcal{A}} = \lambda_{S_{(i_1)}R_m} + \cdots + \lambda_{S_{(i_K)}R_m} + \lambda_{R_mD}. \quad (44)$$

The total system's outage probability of the proposed relay and UE association is dependent on the outage events of direct UE and receiver connections and multihop connections between UE and the associated relay nodes. In this formulation, $\varepsilon_\eta$ and $\mathcal{V}_\ell$ show the set of non-outage UEs and associated relay nodes with degrees $\eta$ and $l$. the close-form expression of $\mathcal{V}_\ell$ and $\varepsilon_\eta$ is shown as follows.

21

$$\varepsilon_\eta \triangleq \{S_{(k)} \in S: \gamma_{S_{(k)}D} > \gamma_{th}\}, \quad (45)$$

$$\mathcal{V}_\ell|\mathcal{A} \triangleq \{R_{(l)} \in \mathcal{R}: min_{\gamma_{(l)|\mathcal{A}}} > \gamma_{th}\}. \quad (46)$$

In which $\gamma_{th}$ is equal to the threshold of acceptable signal to noise ratio. The total outage events of the relay and UE association can then be exhibited as

$$\mathcal{O} = \acute{\mathcal{O}} \cup \grave{\mathcal{O}}, \quad (47)$$

In which $\acute{\mathcal{O}}$ represents the outage occurrence for $K > l$ considering UEs without temporal outage as $\eta$, even if $\ell = L$, the final receiver is in temporal failure i.e., $\eta < K - L$. we can express that, $\grave{\mathcal{O}}$ is related to the outage events where $\eta \geq K - L$, in such a way that the quantity of selected UEs and relay nodes without outage is less than $K$ and we have, $\eta + \ell < K$.

In this scenario, $\gamma_{th}$ indicates the threshold of acceptable SINR, which we can determine its value in accordance with the transmission rate $R_0$ (bits/channel) as $\gamma_{th} = 2^{R_0} - 1$. If $\mathcal{O}_{ij}$ is the outage occurrence of connection $i \rightarrow j$. The outage probability of $i \rightarrow j$ connection will be expressed as (48)

$$Pr\{\mathcal{O}_{ij}\} = Pr\{\gamma_{ij} < \gamma_{th}\}, \quad (48)$$

In which $\gamma_{ij} = \rho|h_{ij}|^2$ is the instantaneous signal to noise ratio of connection $i \rightarrow j$. It should be noted that $\gamma_{ij}$ has exponential distribution, and (48) is properly solved as

$$Pr\{\mathcal{O}_{ij}\} = 1 - \int_{\gamma_{th}}^{\infty} \lambda_{ij} e^{-\lambda_{ij}y} dy = 1 - e^{-\lambda_{ij}\gamma_{th}}, \quad (49)$$

Considering statistics order and applying (49), we can express $Pr\{\varepsilon_\eta\}$ as (40) and the close-form expression of $Pr\{\mathcal{V}_\ell\}$ is obtained via (50) through total probability assumptions.

$$Pr\{\mathcal{V}_\ell\} = \sum_{\mathcal{A}} Pr\{\mathcal{V}_\ell|\mathcal{A}\} Pr\{\mathcal{A}\}, \quad (50)$$

To obtain the optimal value of $Pr\{\mathcal{V}_\ell|\mathcal{A}\}$ via (39), all $\binom{N}{K}$ possibilities is considered which $N$ is the set of candidate UEs.

$$Pr\{\mathcal{V}_\ell|\mathcal{A}\} = \sum_{v=M-j_{\ell+1}+1}^{M-j_\ell} \left( \sum_{\substack{a_1,\ldots,a_v \in \{1,\ldots,M\} \\ a_1 \neq \cdots \neq a_v}} \left[ \prod_{m=a_1}^{a_v} Pr\{\mathcal{O}_m|\mathcal{A}\} \prod_{\substack{\acute{m}=1 \\ \acute{m} \neq \{a_1,\ldots,a_v\}}}^{M} (1 - Pr\{\mathcal{O}_{\acute{m}}|\mathcal{A}\}) \right] \right). \quad (51)$$

Also, $Pr\{\mathcal{A}\}$ is obtained by (52) in which all $(N-K)!K!$ possible cases should be considered. For instance, if $\lambda_{S_nD} \approx \lambda_{SD}, \forall n$, $Pr\{\mathcal{A}\}$ is equal to $Pr\{\mathcal{A}\} \approx \frac{1}{\binom{N}{K}}$.

$$Pr\{\mathcal{A}\} = \sum_{\substack{z_1,\ldots,z_N \in \{1,\ldots,N\} \\ z_1 \neq \cdots \neq z_N \\ z_{i_1},\ldots,z_{i_K} \in \mathcal{A}}} \left( \prod_{n=2}^{N} \left[ \frac{\lambda_{S_{z_n}D}}{\lambda_{S_{z_1}D} + \sum_{i=2}^{n} \lambda_{S_{z_i}D}} \right] \right) \quad (52)$$

Based on the above-mentioned formulation, considering (50), (51) and (52), the $Pr\{\mathcal{V}_\ell\}$ can be expressed as (41). Also, taking (40), (41), and (47) into account, the total outage probabilities calculated through (38) and (39).

## 5. NUMERICAL RESULTS

In this section, we utilize MATLAB simulations to assess the performance of the proposed algorithm, Energy Efficient Caching-enabled MEC (EE-CMEC), and evaluate it compared to other methodologies such as Fixed Power Allocation (FPA), Random Power Allocation (RPA), and Cooperative NOMA Simultaneous Wireless Information and Power Transfer (CN-SWIPT) [57] in two modes: 1) with power sharing (PS) capability and 2) without power sharing. In the case of fixed power allocation, all base stations supply the same maximal transmission power to user equipment and backhaul connections, so each backhaul connection and user equipment use all available resources. The random power allocation follows each t in C1 through C5, regardless of the objective function. In addition to the deep learning-based UA and RA algorithms described in a separate mathematical model, the deep learning technique can effectively seek the ideal network design, delivering the best continuing power flow with the largest possible load limit. Multiple deep learning methods are used to forecast the output parameters. Table 1 includes a list of the key simulation variables as well as their principal quantities. Figures 5 and 6 compare the average total throughput of the EE-CMEC method to various EE systems when the highest transmission power and UE densities are identical. Figure 5 depicts the simulated results depending on the uniform user distribution pattern, whereas Figure 6 depicts the user throughput depending on the mixed user distribution pattern (hotspot and mobile). To these results, EE-CMEC performs significantly better than other algorithms in terms of overall network throughput. Because of its adaptability to time-varying fade variables, this method offers the essential flexibility to deliver increased output for user equipment. This output may also fall significantly as *N* increases, because the method limits the output accessible to each UE to ensure the fairness ratio. The structure of user requests, on the other hand, is consistent across all randomized power allocation, constant power allocation, and EE-CMEC algorithms since they all give the smallest output for each *N* user group. As a consequence, the combination of random power allocation and discrete power control beats constant power allocation, yet both can deliver much less output for UEs than CN-SWIPT and EE-CMEC due to inefficiencies and smart user association. According to these figures, the overall network output of the EE-CMEC algorithm improves roughly linearly with increasing *N*. whereas the other three methods show very minor improvements.

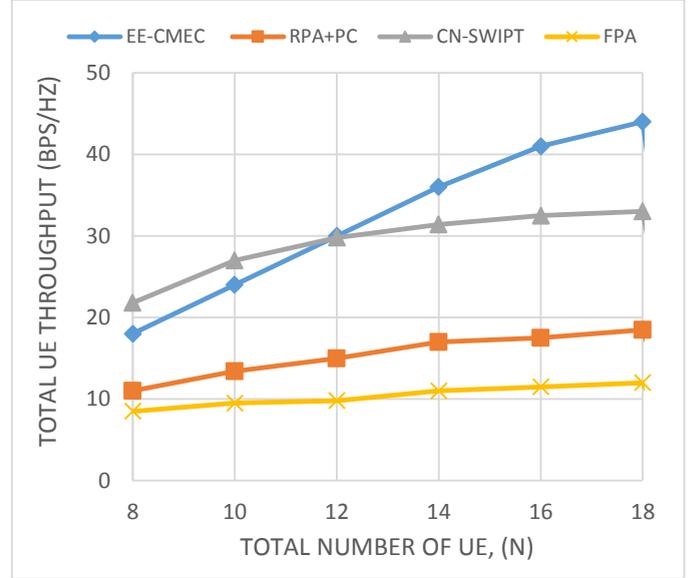

**Fig. 6.** Overall output using hotspot and mobile user distribution models.

Figure 7 compares the average overall power savings of several methods for various user counts. This chart

**Table 1.** The Main Simulation Parameters

| Parameter | Value |
|---|---|
| HetNet Configuration | Multi-Rat HetNet (NGMN), 802.11 |
| Small cell distribution pattern | uniform (U) and hotspot (Hs), |
| Operational BW (MHz) | 2*20 Megahertz |
| Power backoff | 3 dB |
| Max transmit power of macro-BS | 43 dBm |
| $C_h$ | 1 |
| Inter site spacing | 250 m |
| Hopping method | Synthesized frequency hopping |
| $loss\ R_x$ & $loss\ T_x$ | 5 dB |
| Blocking probability | $P_{rt} = 0.4$ |
| Scheduler | Fair |
| Maximum allowed iteration | 500 |
| $L_{margin}$ | 13 dBm |
| Operational FREQ of BH link | 6 GHz |
| Maximum weighting factor $\omega_{max}$ | 0.95 |
| Average session duration | $1/v = 1.5$ |
| Power consumption of macro-BS $P_m^o$ | 60 w |
| Power consumption of small-BS $P_s^o$ | 1.5 w |
| $\$_{VHO}$ & $\$_C$ & $\$_w$ | 6.4 & 10 & 0.5 |
| $P_{hotspot}$ | 0.5 |

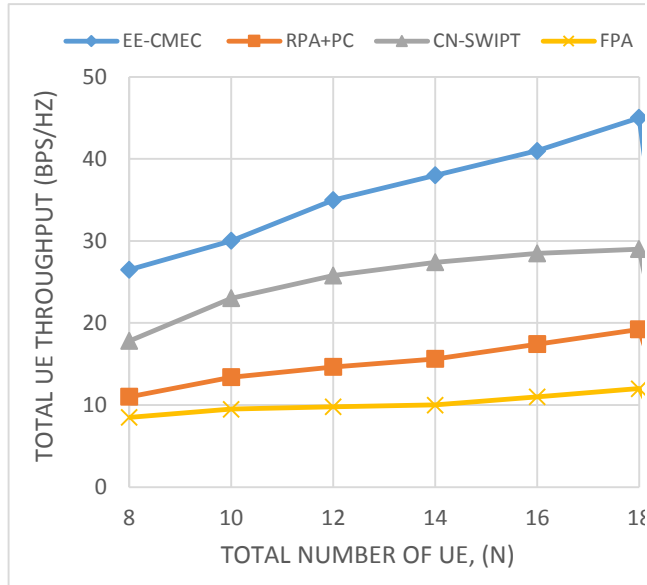

**Fig.5.** shows the total UE output when the uniform user distribution approach is used.

shows that the energy savings of the CN-SWIPT and EE-CMEC algorithms is significantly higher than that of randomized power allocation and constant power allocation. Therefore, in addition to dynamic user association, the CN-SWIPT and EE-CMEC algorithms maximize power allocation and flow control. Nevertheless, due to variations in their target functions,

23

EE-CMEC outperforms CN-SWIPT. As previously stated, the main function of EE-CMEC is to optimize energy savings. To that end, it improves the requirements for user equipment performance as well as energy usage in both the access network and backhauling. According to the achieved graph, the network's output weight is greater than the rise in power usage for the N UE set, which promotes energy savings while raising the N with a steep gradient. However, once N grows greatly, energy saving capacity stays unchanged since increasing capacity is ineffective owing to rising energy usage. Because network capacity cannot maintain the required balancing in high energy usage, energy savings may deteriorate progressively as N increases. However, CN-SWIPT simply improves energy usage, whereas UE needs to face major technical constraints. This demonstrates that, in terms of energy saving, it is preferable to improve UE performance needs, energy usage, and control current in backhaul lines at the same time.

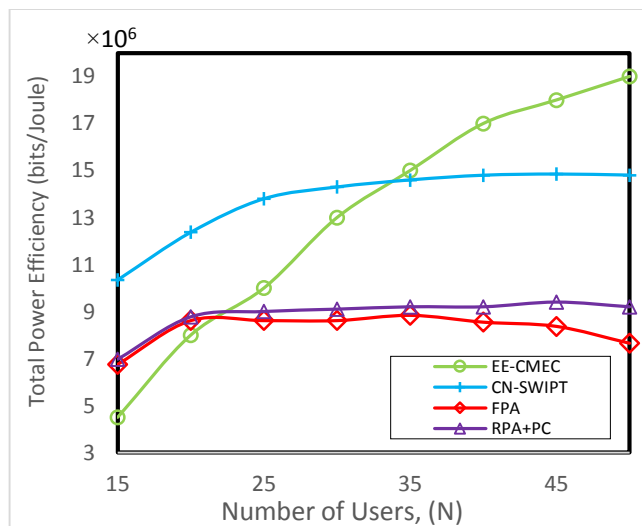

**Fig.7.** total power saving vs the quantity of UEs.

As expected, Fig. 7 indicates that the EE-CMEC algorithm makes greater use of power sources than other algorithms. Raising the highest transmission power to saturation improves energy saving. Raising the highest transmission power after achieving the saturation level has no effect on energy saving. Hence, raising the highest transmission power in Fixed Power Allocation (FPA), Random Allocation (RA), and CN-SWIPT algorithms does not enhance energy saving. The efficiency of these algorithms is marginally inferior in this scenario. We studied how many times each method efficiently obtained the ideal answer based on the given parameters in CVX using 1000 separate simulations.

## 6. CONCLUSION

Within that study, we look at how to maximize energy savings in mobile edge computing networks with various backhaul link connections. To discover the best architecture for cache-enabled backhaul networks, we presented a DL-based allocation of resources technique. A variety of radial layouts of test equipment have been used in practice settings for training phases. This article also used the continuous power flow (CPF) assessment to calculate the optimal load limit wherein the power of macro base stations with varying cache capacities is supplied by renewable energy systems or the smart grid. Power sharing capability was allowed across multiple layers of network elements using smart grids to boost the power saving index of this strategy. We have shown that this approach allows cluster formation by taking into account the position of nodes and their neighbors at the same time. Our experimental results showed that this method of channel allocation compared to other methods significantly reduces internal and external interference and improves the saturation and overall throughput in the network. The quantitative experiments demonstrated that the proposed method can achieve improved energy savings, overall data rate, and user fairness. It also beats other methods which do not consider the general influence on overall data rate maximization and energy efficiency. This study is presently undertaken to create a low-complexity dynamic process of implementing the EE-CMEC paradigm.